\documentclass[amssymb, nobibnotes, aps,twocolumn]{revtex4-1}
\usepackage{graphicx}

\usepackage{amsmath}
\usepackage{times}

\def\cc{\rm c.c.}
\def\hc{\rm H.c.}
\def\sub#1{_{\rm #1}}
\def\ee{{\rm e}}
\def\ii{{\rm i}}
\def\ket#1{|#1\rangle}
\def\bra#1{\langle #1|}
\def\dd{{\rm d}}
\def\U#1{{%
\def\O{\mbox{O}}
\def\u{\mbox{u}}
\mathcode`\u=\mu
\mathcode`\O=\Omega\rm #1}}

\begin{document}

\title{Implementation of Electromagnetically Induced
Transparency \\
in a Metamaterial Controlled with Auxiliary Waves}

\author{Toshihiro Nakanishi}
\email[E-mail: ]{t-naka@kuee.kyoto-u.ac.jp}
\author{Masao Kitano}
\affiliation{Department of Electronic Science and Engineering,
Kyoto University, Kyoto 615-8510, Japan}
\date{\today}%

\begin{abstract}
We propose a metamaterial to realize true electromagnetically induced transparency (EIT), where the incidence of an auxiliary electromagnetic wave called the control wave induces transparency for a probe wave.
The analogy to the original EIT effect in an atomic medium is shown through analytical and numerical calculations
derived from a circuit model for the metamaterial.
We perform experiments to demonstrate the EIT effect of the metamaterial in the microwave region.
The width and position of the transparent region can be controlled by the power and frequency of the control wave.
We also observe asymmetric transmission spectra unique to the Fano resonance.

\end{abstract}


\maketitle
%\tableofcontents

\section{Introduction}

Electromagnetically induced transparency (EIT) is a nonlinear optical effect that renders an opaque medium
transparent in a narrow spectral region due to the incidence of auxiliary light called control light \cite{Harris1997, Fleischhauer2005}.
The EIT effect is caused by destructive interference between two excitation pathways in three-level atoms.
In addition to the absorption, the refractive index of the medium is significantly modified in the transparency region, and the group velocity of the light is dramatically reduced \cite{Hau1999}.
The linewidth of the transparency region and the group velocity can be controlled by the intensity of the control light.
The slow-light effect in the EIT medium was developed to store light; this is realized by temporal control of the EIT effect through switching of the control light.
The storage of light has been widely studied to realize optical memory and quantum memory \cite{Phillips2001, Liu2001,Turukhin2002}.

The interference phenomenon is found not only in quantum systems but
also in classical systems, and classical analogs of the atomic EIT
effects are demonstrated in various systems, such as the classical
coupled oscillator \cite{GarridoAlzar2002}, optical waveguide coupled
with cavities \cite{Little1999,Xu2006,Totsuka2007,Kekatpure2010},
optomechanical system\cite{Lin2009a, Weis10122010, Safavi-Naeini2011}, 
and acoustic system \cite{Yoo2014}.
In addition, there are many ways to mimic atomic EIT effects with electromagnetic metamaterials, which are assemblies of artificial structures (i.e., meta-atoms), at a scale much smaller than the operating wavelength.
Since experimental demonstrations in the microwave region were first performed \cite{Fedotov2007,Papasimakis2008},
the sharp transparency and resulting slow propagation have been demonstrated at higher frequencies, including the terahertz \cite{Chiam2009,Li2011,Singh2011,Liu2012} and optical ranges \cite{Liu2009,Zhang2010,Huang2011a,Hokari2014,Yang2014}.

The operation of these EIT-like metamaterials is well described by a coupled-resonator model composed of a high-loss resonator interacting with a propagating wave and a low-loss resonator decoupled from the wave.
The former resonance mode is referred to as the {\it radiative} or {\it bright mode}, and the latter is called the {\it trapped} or {\it dark mode}.
When these two resonators are coupled, the energy received through the radiative mode is transferred into
the trapped mode with a long lifetime, and the dissipation and  propagation speed are substantially
reduced.
If the resonant frequencies of the two resonators are identical, a sharp transparent region appears at the center of a broad absorption profile with a symmetrical shape, like in the original atomic EIT effect.
On the other hand, when the resonant frequencies are different, the transparent region is located at the shoulder of the broad absorption profile and shows an asymmetrical shape \cite{Fedotov2007,Luk'yanchuk2010,Giannini2011}, which is unique to the Fano resonance \cite{Fano1961, Miroshnichenko2010}.
In addition to slow propagation, EIT-like metamaterials are applied to
implementing various functionalities, such as accurate sensing \cite{Lahiri2009,Dong2010,Liu2010a}, manipulation of near fields \cite{Zhang2012}, nonreciprocal transmission \cite{Sun2013,Mousavi2014}, lasing spacers \cite{Zheludev2008}, and absorption enhancement \cite{Taubert2012}.

In order to realize practical applications, including the storage of electromagnetic waves, the tunability of the EIT-like effects is quite important.
Many researchers report various methods to tune EIT-like metamaterials in a passive manner by changing the incident angles \cite{Tamayama2012, Jin2012} and in active manners by conductivity modulation utilizing
diodes \cite{Meng2014b},  superconductors \cite{Kurter2011, Limaj2014}, and photocarrier excitation in a semiconductor \cite{Gu2012, Miyamaru2014} or by tuning an external magnetic field \cite{Mousavi2014}.
Recently, we proposed an EIT-like metamaterial whose properties can be
controlled by applying bias voltages to diodes to change their
capacitances, and we experimentally demonstrated the storage of
electromagnetic waves in the microwave region \cite{Nakanishi2013}.
This method can be regarded as static-electric-field-induced transparency and is promising owing to the tunability. However, the static field cannot propagate in free space, and it should be individually fed to each meta-atom through a bias circuit, which complicates the structure and causes difficulty in increasing the number of meta-atoms.

The metamaterials introduced above reproduce the sharp transparency and slow propagation unique to the atomic EIT effect; however, the analogy is incomplete in the sense that there is no counterpart to the control light, which plays an important role in controlling the EIT effect in the atomic system.
In this sense, it is not appropriate to use "{\it electromagnetically
induced transparency}" to refer to these metamaterials.
In this paper, we propose a method to implement true EIT in a metamaterial; in other words, the functions of the metamaterial can be controlled by the incidence of an auxiliary electromagnetic wave or control wave.
The proposed metamaterial also has a radiative mode and a trapped mode
whose resonant frequencies $\omega\sub{r}$ and $\omega\sub{t}$, respectively, are quite different.
If nonlinear elements are properly installed in the metamaterial, the two resonant modes can be coupled in the presence of the control wave oscillating at $|\omega\sub{r}-\omega\sub{t}|$ through the parametric process.
The coupling mechanism, which we call nonlinearity-assisted coupling, is employed in several studies
to realize an effective magnetic field for photons \cite{Fang2012,Fang2012b,Fang2013} or to enhance second-harmonic generation in a metamaterial \cite{Nakanishi2012b}.
We can show that the time evolution of the coherence of three-level atoms in the atomic EIT medium and the charge oscillation in our metamaterial are governed by the same differential equations.
As a result, the electric susceptibilities of the atomic EIT medium and our metamaterial can be written with the same form, which means that it is impossible to differentiate the metamaterial from the atomic EIT medium simply by measuring the electromagnetic response in an effective-medium approximation.

In this paper, we first review the EIT effect in an atomic medium in Sec.~\ref{sec:atom} and introduce the metamaterial loaded with nonlinear capacitors to implement the true EIT effect in Sec.~\ref{sec:model}.
We explicitly describe the analogy to the atomic EIT effect and compare the original EIT system to the circuit model of the metamaterial.
For both systems, we define different physical quantities with the same notations to explicitly express the relationship.
In Sec.\ref{sec:simulation}, we examine the validity of the approximated expressions derived in Sec.~\ref{sec:model} through numerical calculations and describe the physical mechanism of the
EIT effect in the metamaterial analog.
In Sec.~\ref{sec:design}, 
we present the actual design of the metamaterial, which has a resonant mode for a control wave besides the radiative mode and the trapped mode.
We show the linear response of the metamaterial calculated by electromagnetic simulation and discuss the role of each resonant mode.
In Sec.~\ref{sec:exp}, we demonstrate the control of the EIT effect by using a control wave in the microwave region.
We can control the width of the transparency window by changing the intensity of the control wave.
Furthermore, we can control the position of the transmission peak and the Fano line shape by tuning the frequency of the control wave.

Compared with the static-electric-field-induced transparency\cite{Nakanishi2013}, which requires the bias circuit on each meta-atom, the functions to receive the control wave and modulate the capacitances for frequency mixing are integrated into the metamaterial. We can tune the EIT effect just by illuminating the metamaterial with the propagating control wave. In addition, the resonance enhancement for the control wave is also one of the advantages, and we can enhance the efficiency of the control wave by increasing the quality factor of the resonance mode for the control wave.

\section{Atomic EIT effect and metamaterial analog\label{sec:theory}}

\subsection{Atomic EIT\label{sec:atom}}
\begin{figure}[b]
 \begin{center}
 \includegraphics[scale=0.5]{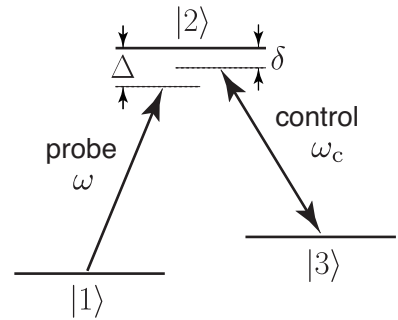}
 \caption{The atomic EIT system.}
 \label{fig:atom}
 \end{center}
\end{figure}

Here, we review the original EIT effect in a quantum system composed of three-level atoms, as shown in Fig.~\ref{fig:atom}, before introducing a metamaterial analog.
If a probe light with the electric field $E \ee^{-\ii \omega t}+\cc$, where $\cc$ represents a complex conjugate, and a control light with the electric field $E\sub{c} \ee^{-\ii(\omega\sub{c} t+\phi)}+\cc$ induce the electric-dipole transitions $\ket{1} \rightarrow \ket{2}$ and $\ket{3}\rightarrow \ket{2}$, respectively, the Hamiltonian of the three-level system interacting with these electric fields in a rotating wave approximation is expressed as
\begin{align}
 H = &\hbar \{\omega\sub{1} \ket{1}\bra{1}
 + \omega\sub{2} \ket{2}\bra{2}
 + \omega\sub{3} \ket{3}\bra{3}\} \nonumber \\
 &- \hbar \{ \Omega \ket{2}\bra{1} \ee^{-\ii \omega t} +
 \Omega\sub{c} \ket{2}\bra{3} \ee^{-\ii (\omega\sub{c} t+\phi)} + \hc\},
\end{align}
where $\hc$ represents Hermitian conjugate terms and the Rabi frequencies are defined as $\Omega=pE/\hbar$ and
$\Omega\sub{c}=p\sub{c}E\sub{c}/\hbar$ with the electric-dipole moments
$p$ and $p\sub{c}$ of the transitions for the probe and control light, respectively.
The evolution of the density matrix $\rho$ can be calculated by
\begin{align}
 \ii \hbar \frac{\dd \rho}{\dd t} = [ H, \rho ].
\end{align}
The control light is strong enough to populate only $\ket{1}$, and we can fairly assume $\rho_{11} \approx 1, \rho_{22}=\rho_{33}=\rho_{23}\approx 0$.
As a result, the two matrix elements for coherence $\rho_{21}$ and $\rho_{31}$ should be calculated.
By introducing rotating frames defined as $\rho_{21}=\tilde{\rho}_{21} \ee^{-\ii \omega t}$ and $\rho_{31}=\tilde{\rho}_{31} \ee^{-\ii (\omega-\omega\sub{c}) t}$, the set of equations can be obtained as
\begin{align}
 \frac{\dd \tilde{\rho}_{21}}{\dd t} &= -(\gamma+\ii \Delta)
 \tilde{\rho}_{21} + \ii \Omega + \ii \Omega\sub{c} \ee^{-\ii \phi}
 \tilde{\rho}_{31}, \label{r21}\\
 \frac{\dd \tilde{\rho}_{31}}{\dd t} &= -\{\gamma\sub{t} +\ii
 (\Delta-\delta) \}
 \tilde{\rho}_{31} + \ii \Omega\sub{c} \ee^{\ii \phi} \tilde{\rho}_{21},\label{r31}
\end{align}
where the detunings are defined as $\Delta=\omega_2-\omega_1-\omega$ and $\delta=\omega_2-\omega_3-\omega\sub{c}$.
In the derivation, we can phenomenologically introduce $\gamma$ and $\gamma\sub{t}$ for the relaxations $\ket{2} \rightarrow \ket{1}$ and $\ket{3} \rightarrow \ket{1}$, respectively.
We can assume $\gamma\gg\gamma\sub{t}$ because the relaxation rate between the ground states is much slower than that from the excited state.
In the steady state, the coherence between $\ket{1}$ and $\ket{2}$ can be derived as
\begin{align}
 \tilde{\rho}_{21} = \frac{\ii \Omega\{\gamma\sub{t} +\ii (\Delta-\delta)\}}{
 (\gamma+\ii \Delta) \{ \gamma\sub{t} + \ii (\Delta-\delta)\} + \Omega\sub{c}^2
 }.
\end{align}
The complex susceptibility of the medium composed of three-level atoms is proportional to the coherence $\tilde{\rho}_{21}$ because $\chi\sub{e}=(N p \tilde{\rho}_{21})/(\epsilon_0 E)$, where $N$ is the density of the atoms.
Thus, the susceptibility is given as follows \cite{Scully,Agarwal}:
\begin{align}
 \chi\sub{e} = \frac{\ii p^2 N}{ \epsilon_0 \hbar}
 \frac{\gamma\sub{t} +\ii (\Delta-\delta)}{
 (\gamma+\ii \Delta) \{ \gamma\sub{t} + \ii (\Delta-\delta)\} + \Omega\sub{c}^2
 }. \label{chi_atom}
\end{align}
The susceptibility spectra are considered in detail in
Sec.~\ref{sec:simulation} through the use of the metamaterial analog described in
Sec.~\ref{sec:model}.

\subsection{Metamaterial analog and circuit model\label{sec:model}}

\begin{figure}[b]
 \begin{center}
 \includegraphics[scale=0.43]{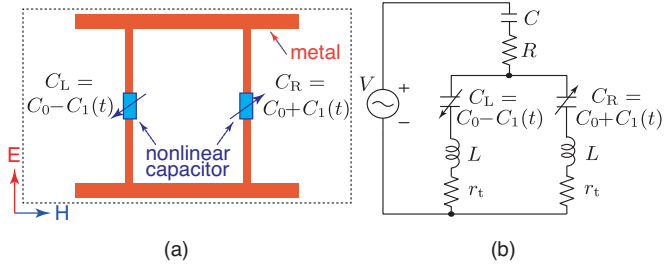}
 \caption{(a) The schematics of the unit cell. (b) The circuit model.}
 \label{fig:unit}
 \end{center}
\end{figure}

Here, we introduce a metamaterial analog to the atomic EIT medium.
The unit cell is a metallic structure loaded with the two nonlinear capacitors $C\sub{R}$ and $C\sub{L}$ on two arms, as shown in Fig.~\ref{fig:unit}(a).
We derive the electromagnetic response for a probe wave to be observed with the electric field ${E}$ and magnetic field ${H}$, as depicted in Fig.~\ref{fig:unit}(a), when an auxiliary wave (i.e., {\it control wave}) is also incident to the metamaterial in order to modulate the nonlinear capacitances as $C\sub{R}=C_0+C_1(t)$ and $C\sub{L}=C_0-C_1(t)$.
Here, $C_1(t)=C\sub{m} \cos(\omega\sub{c}t+\phi)$.
This is discussed in detail in Sec.~\ref{sec:design}.
This metamaterial can be well described by a circuit model, as illustrated in Fig.~\ref{fig:unit}(b).
The circuit elements $C$ and $L$ represent the capacitance between neighboring unit structures in the vertical direction and the inductance of each metallic arm, respectively.
The source voltage $V$ is determined by the external electric field $E$ as $V=El$, where $l$ is the height of the unit cell.
The physical meanings of the resistors $R$ and $r\sub{t}$ are discussed later.
With this circuit model, the equations of motion for $q_{\pm} \equiv q\sub{R} \pm q\sub{L}$, where $q\sub{R}$ and $q\sub{L}$ are the charges of $C\sub{R}$ and $C\sub{L}$, respectively, can be written as follows:
\begin{align}
 &L \frac{\dd^2 q_+}{\dd t^2} + r \frac{\dd q_+}{\dd t} +
 \frac{q_+}{C'} - \frac{C\sub{m}}{C\sub{0}^2} \cos(\omega\sub{c}t+\phi) q_- = 2V, \label{ori3}\\
 &L \frac{\dd^2 q_-}{\dd t^2} + r\sub{t} \frac{\dd q_-}{\dd t} +
 \frac{q_-}{C_0} - \frac{C\sub{m}}{C\sub{0}^2} \cos(\omega\sub{c}t+\phi) q_+ = 0,
 \label{ori4}
\end{align}
where we suppose $V=V_0\cos\omega t$ and define
$r=2R+r\sub{t}$ and $1/C'=2/C+1/C_0$.
In this derivation, we assume the small modulation $C_0 \gg C\sub{m}$.

Without modulation (i.e., for $C\sub{m}=0$), Eq.~(\ref{ori3}) represents a harmonic oscillator for $q_+$ with the resonant frequency $\omega\sub{r}=1/\sqrt{L C'}$ driven by the external source $2V_0 \cos\omega t$.
Equation (\ref{ori4}) also represents another harmonic oscillator for $q_-$ with the different resonant frequency $\omega\sub{t}=1/\sqrt{L C_0}$.
The latter oscillator is uncoupled from the probe wave.
The resistances $r$ and $r\sub{t}$ are the radiation resistances for the
in-phase charge ($q_+$) oscillation and out-of-phase charge ($q_-$)
oscillation, respectively.
We can assume $r \gg r\sub{t}$, because the radiation from the electric-dipole oscillation due to $q_+$ is much greater than that from the magnetic-dipole oscillation due to $q_-$.
The oscillation of $q_+$ works as a radiative mode that receives and emits electromagnetic waves, and the oscillation of $q_-$ works as a trapped mode that temporarily stores the electromagnetic energy with low losses.

On the other hand, when  $C\sub{m} \neq 0$, the two resonant modes with different resonant frequencies $\omega\sub{r}$ and $\omega\sub{t}$ are effectively coupled if the modulation frequency $\omega\sub{c}$ satisfies $\omega\sub{c} \approx \omega\sub{r}-\omega\sub{t}$.
The coupling between a lossy resonator interacting with an external
field and a low-loss resonator is a requirement for realizing the EIT effect in this model.
This fact can be confirmed by solving Eqs.~(\ref{ori3}) and (\ref{ori4}) under the near-resonance conditions $|\omega-\omega\sub{r}| \ll \gamma$ and $|(\omega-\omega\sub{c})-\omega\sub{t}| \ll \gamma\sub{t}$, where the relaxation rates of the resonant modes are defined as $\gamma=r/(2L)$ and $\gamma\sub{t}=r\sub{t}/(2L)$.
Under these conditions, $q_\pm$ can be written in the forms of $q_+=\tilde{q}_+ \ee^{-\ii \omega t}+\cc$ and
$q_-=\tilde{q}_- \ee^{-\ii (\omega-\omega\sub{c}) t}+\cc$, where
$\tilde{q}_\pm$ represents slowly varying envelopes for $q_\pm$, because
other frequency components produced by the mixing processes are off resonance.
Then, the following equations are obtained:
\begin{align}
 \frac{\dd \tilde{q}_+}{\dd t} &= -(\gamma+\ii \Delta)
 \tilde{q}_+ + \ii \frac{V_0}{2 \omega\sub{r} L}
 + \ii \frac{\omega\sub{t}^2}{4 \omega\sub{r}} \frac{C\sub{m}}{C_0}
 \ee^{-\ii \phi}\tilde{q}_{-}, \label{q+}\\
 \frac{\dd \tilde{q}_-}{\dd t} &= -\{\gamma\sub{t} +\ii
 (\Delta-\delta) \}
 \tilde{q}_{-} + \ii \frac{\omega\sub{t}}{4}
 \frac{C\sub{m}}{C_0}
 \ee^{\ii \phi} \tilde{q}_{+}, \label{q-}
\end{align}
where $\Delta=\omega\sub{r}-\omega$ and $\delta=\omega\sub{r}-\omega\sub{t}-\omega\sub{c}$.
We assume that $C\sub{m} \ll C_0$ and $\omega+\omega\sub{r} \sim 2 \omega\sub{r}$.
As a result, the circuit model is reduced to a coupled oscillator that can be compared with the atomic EIT system
governed by Eqs.~(\ref{r21}) and (\ref{r31}).
The coherent oscillation between the excited state $\ket{2}$ and one of the ground states $\ket{1}$ in the atomic system is represented by the in-phase oscillation $q_+$, which forms an electric-dipole oscillation that interacts with the external field.
On the other hand, the coherent oscillation between the two ground states with a long relaxation time is represented by the out-of-phase oscillation $q_-$ with low losses.
The coupling coefficients seem to differ from each other in the last terms of Eqs.~(\ref{q+}) and (\ref{q-}),
but these equations can be rewritten in the exact same form as Eqs.~(\ref{r21}) and (\ref{r31}) by introducing the new variable $\tilde{q}_-'=\sqrt{\omega\sub{t}/\omega\sub{r}}\, \tilde{q}_-$ and a coupling coefficient:
\begin{align}
 \Omega\sub{c} = \frac{\omega\sub{t}}{4}
 \sqrt{\frac{\omega\sub{t}}{\omega\sub{r}}} \frac{C\sub{m}}{C_0}.
\end{align}
In the EIT metamaterial, $\Omega\sub{c}$ is proportional to the modulation amplitude $C\sub{m}$ or the electric field of the control wave, as in the case of the original atomic system.

The electromagnetic wave interacts only with the in-phase mode $q_+$, which forms the electric-dipole oscillation.
The electromagnetic response of the metamaterial is determined by the electric susceptibility $\chi\sub{e}=(N \tilde{p})/(\epsilon_0 E)$, where $N$ is the density of the unit structures and $\tilde{p}$ is the electric-dipole moment induced in a single structure.
By using $\tilde{p}=\tilde{q}_+ d$, where $d$ is the effective dipole length for the in-phase oscillation, we obtain
\begin{align}
 \chi\sub{e} =
 \ii \frac{N d l}{2 \epsilon_0 \omega\sub{r} L}
\frac{ \{ \gamma\sub{t} +\ii (\Delta-\delta) \} }{
 (\gamma+\ii \Delta) \{ \gamma\sub{t} +
 \ii (\Delta-\delta)\} + \Omega\sub{c}^2
 }. \label{chi}
\end{align}
This susceptibility and that of the atomic EIT medium given by Eq.~(\ref{chi_atom}) are the same except for the
coefficients.
This fact guarantees that the artificial medium with this metamaterial responds to the probe wave in the exact same way as the original EIT system composed of the three-level atoms.

\subsection{Numerical simulation of circuit model\label{sec:simulation}}

\begin{figure}[b]
 \begin{center}
 \includegraphics[scale=0.45]{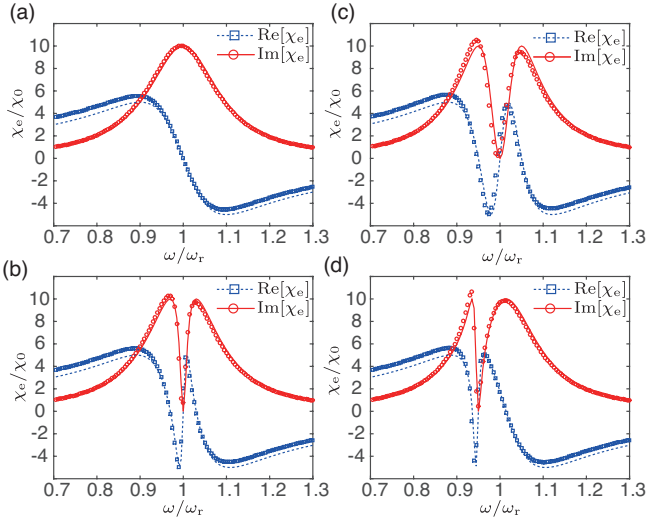}
 \caption{Complex electric susceptibility for
 (a) $\omega\sub{c}=0.3\omega\sub{r}$, $\Omega\sub{c}=0$,
 (b) $\omega\sub{c}=0.3\omega\sub{r}$, $\Omega\sub{c}=0.03\omega\sub{r}$,
 (c) $\omega\sub{c}=0.3\omega\sub{r}$,
  $\Omega\sub{c}=0.05\omega\sub{r}$, and
 (d) $\omega\sub{c}=0.25\omega\sub{r}$, $\Omega\sub{c}=0.03\omega\sub{r}$.
 The normalization constant is defined as $\chi_0=(N d l)/(2 \epsilon_0 \omega\sub{r}^2 L)$.}
 \label{fig:comp}
 \end{center}
\end{figure}

In the previous section, we provide the analytical expressions for the behavior of the EIT metamaterial from Eqs.~(\ref{ori3}) and (\ref{ori4}) by using several approximations such as the near-resonance condition and weak nonlinearity.
Now, we compare the approximated solution given by Eq.~(\ref{chi}) and the numerical solution derived from Eqs.~(\ref{ori3}) and (\ref{ori4}).
The numerical computation is conducted with the help of the differential equation solver of the commercial software \textsc{MATLAB}.

Figures \ref{fig:comp}(a)--\ref{fig:comp}(c) show complex electric susceptibilities for  $\Omega\sub{c}=\{0,\, 0.03\omega\sub{r},\, 0.05\omega\sub{r}\}$, $\delta=0$, $\gamma=0.1\omega\sub{r}$,
$\gamma\sub{t}=0$, $\omega\sub{t}=0.7\omega\sub{r}$, and $\omega\sub{c}=0.3\omega\sub{r}$.
The real part ${\rm Re}[\chi]$ and imaginary part ${\rm Im}[\chi]$ are represented by squares and circles, respectively.
The dotted and solid lines of each graph show ${\rm Re}[\chi]$ and ${\rm Im}[\chi]$ as obtained by Eq.~(\ref{chi}).
Without the capacitance modulation (i.e., $\Omega\sub{c}=0$), the imaginary part ${\rm Im}[\chi]$ concerned with absorption shows a Lorenzian profile, as shown in Fig.~\ref{fig:comp}(a), because $q_+$ and $q_-$ are decoupled and only the $q_+$ mode is excited.
On the other hand, in the presence of modulation, the spectra of ${\rm Im}[\chi]$ exhibit sharp depressions at
$\omega=\omega\sub{r}$ in broad Lorenzian profiles, as shown in Figs.~\ref{fig:comp}(b) and \ref{fig:comp}(c).
These characteristics are peculiar to the EIT effect.
For the incident frequency of $\omega=\omega\sub{r}$, the wave received
through the radiative mode is frequency converted to $\omega-\omega\sub{c}$ through the parametric process, and the converted wave is transferred into the trapped mode owing to the resonance $\omega-\omega\sub{c}=\omega\sub{t}$.
As a result, the absorption of the incident wave is significantly suppressed owing to the low losses in the trapped mode.
A larger modulation $C\sub{m} (\propto \Omega\sub{c})$ is confirmed to mean a wider width of the transparency window, just like for the atomic EIT medium.
The slope of ${\rm Re}[\chi]$ is related to the group velocity of the incident wave, and a steep positive slope contributes to a significantly low group velocity.
These characteristics are observed in the transparency windows, so we can expect slow propagation, which is another phenomenon typical of EIT.

Figure \ref{fig:comp}(d) shows the case for $\omega\sub{c}=0.25\omega\sub{r}$ and $\Omega\sub{c}=0.03\omega\sub{r}$.
The other parameters are the same as in the previous cases.
In this case, the center of the transparency window is slightly shifted to $\omega=0.95\omega\sub{r}$, and the spectrum shows an asymmetric shape typical of the Fano resonance.
For $\Omega\sub{c}\neq 0$, the center of the transparency window is always located at $\omega=\omega\sub{t}+\omega\sub{c}$, which is written as $\Delta=\delta$.
The condition $\Delta=\delta$ is called the two-photon resonance condition in the atomic EIT system because the two ground states $\ket{1}$ and $\ket{3}$ in Fig.~\ref{fig:atom} are connected through a two-photon resonance that contributes to the EIT effect.

The curves obtained by Eq.~(\ref{chi}) approximately agree with the numerical results.
The difference between them is slightly large for the large detuning $\Delta=\omega\sub{r}-\omega$ because of the degradation of the near-resonance condition.
The shift in the transparency window is caused by higher-order nonlinear effects, which are ignored in the derivation of Eq.~(\ref{chi}).

\section{Design of metamaterial\label{sec:design}}

\begin{figure}[b]
 \begin{center}
 \includegraphics[scale=0.5]{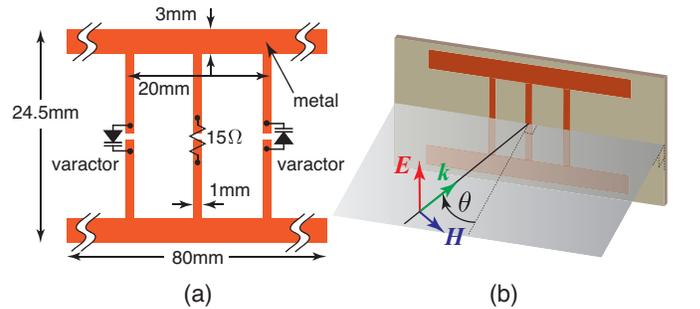}
 \caption{(a) The unit cell structure used in the experiment. (b) The definitions of incident angle and polarization.}
 \label{fig:structure}
 \end{center}
\end{figure}

We implement the EIT metamaterial discussed in the previous section in the microwave region.
The actual design of the unit cell with its dimensions is shown in Fig.~\ref{fig:structure}(a).
Unlike the basic structure shown in Fig.~\ref{fig:unit}(a), the actual unit cell has an additional central bar loaded with a resistor for reasons discussed later.
At the centers of the two outer strips, variable capacitors called varactor diodes are inserted in opposite directions.
The varactor diodes work as nonlinear capacitors because their capacitances are a function of the voltage across the diodes.

To estimate the linear responses for the EIT metamaterial without capacitance modulation, we conduct a numerical simulation using an electromagnetic simulator (\textsc{CST MW Studio}).
We assume that unit structures made of perfect metal are arranged in the
horizontal and vertical directions with the periods of $120\,\U{mm}
\times 25\,\U{mm}$ on a dielectric substrate with a permittivity of 3.3
and a thickness of
0.8\,mm.
The capacitances of the varactor diodes are $C_0=2.5\,\U{pF}$.
The incident angle $\theta$ and polarization are defined in Fig.~\ref{fig:structure}(b).
The incident wave is TE polarized, and the electric field is always aligned in the vertical direction.

\begin{figure}[t]
 \begin{center}
 \includegraphics[scale=0.55]{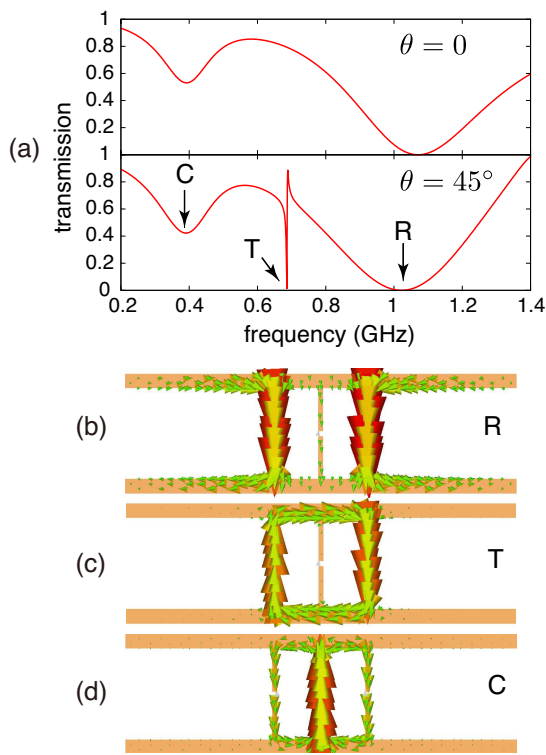}
 \caption{(a) The transmission spectra for $\theta=0$ (upper) and  $\theta=45\,^\circ$ (bottom). Current distributions at (b) $1.02$, (c) $0.687$, and (d) $0.39\,\U{GHz}$ for $\theta=45\,^\circ$.}
 \label{fig:simulation}
 \end{center}
\end{figure}

Figure~\ref{fig:simulation}(a) shows the calculated transmission spectra from $0.2$ to $1.4\,\U{GHz}$
for the normal incidence $\theta=0$ (upper) and oblique incidence $\theta=45^\circ$ (bottom).
The metamaterial has three resonant modes, all of which are observed for the oblique incidence $\theta=45^\circ$.
The positions of these resonances at $1.02$, $0.687$, and $0.39\,\U{GHz}$, are labeled as points R, T, and C, respectively.
The current distributions at points R, T, and C are shown in Figs.~\ref{fig:simulation}(b), \ref{fig:simulation}(c), and \ref{fig:simulation}(d), respectively.
The currents for points R and T mainly flow in the two outer bars, not in the central bar.
The directions of the currents in the outer bars are in phase for point
R and out of phase for point T.
Hence, the resonance at point R can be regarded as the radiative mode, and that at point T can be regarded as the trapped mode.
We also confirm that the linewidth of the trapped mode is quite narrow.
The resonance of the trapped mode cannot be observed for the normal incidence $\theta=0$ because the magnetic flux
does not penetrate the loop forming the trapped mode for the normal incidence.
On the other hand, the other modes are formed by electric-dipole oscillations and are excited at any incident angle.
The small shifts in resonances for various $\theta$ are caused by the mutual coupling among different unit cells.

The resonant mode excited around point C is used to effectively modulate the capacitances of the varactor diodes.
If the control wave tuned around point C is also incident into the metamaterial, the large in-phase voltages are applied to the varactor diodes due to the resonance, as shown in Fig.~\ref{fig:simulation}(d).
Owing to the antisymmetric arrangement of the varactor diodes, each capacitance is modulated in the opposite phase to induce the EIT effect, as discussed in the previous section.
We call this resonant mode the  control mode.
We deliberately insert a resistor in the central bar to widen the tuning range of the control wave, which should be within the linewidth of the control mode.
Note that the amplitude of the capacitance modulation $C\sub{m}$ is proportional to the electric field of the control wave.

\section{Experimental demonstration\label{sec:exp}}

\begin{figure}[t]
 \begin{center}
 \includegraphics[scale=0.8]{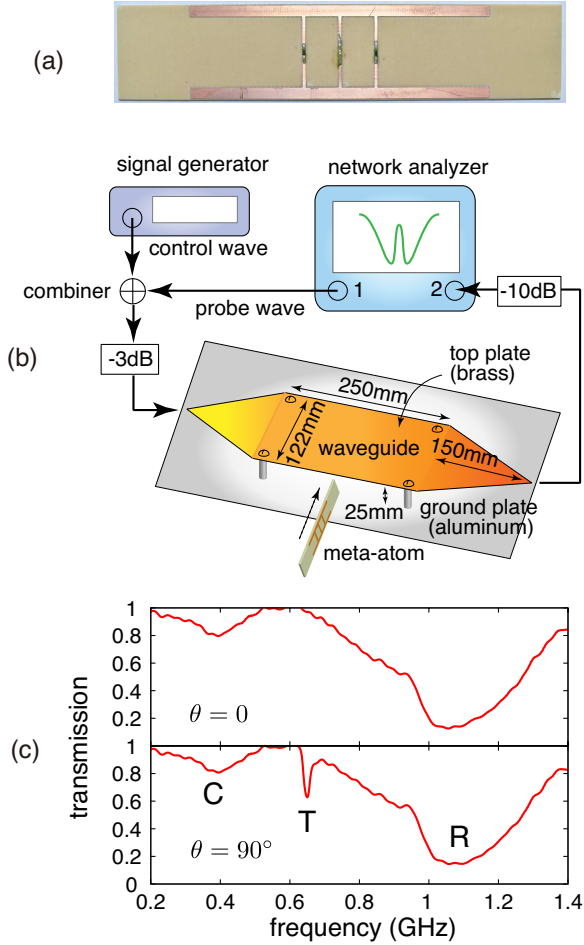}
 \caption{(a) A photograph of a meta-atom. (b) The experimental
  setup. (c) The transmission spectra for normal
 incidence $\theta=0$ (upper) and orthogonal incidence $\theta=90^\circ$ (bottom). }
 \label{fig:setup}
 \end{center}
\end{figure}

We perform an experiment to demonstrate the control of the EIT effect in the metamaterial.
A structure made of copper is fabricated on a print circuit board with dimensions of $120\,\U{mm} \times 25\,\U{mm}
\times 0.8\,\U{mm}$ and a permittivity of 3.3, as shown in Fig.~\ref{fig:setup}(a).
We insert varactor diodes (Infineon BBY52-02W) with $C_0=2.5\,\U{pF}$ at the outer bars and a 15-$\Omega$ resistor at the center bar.
The experimental setup is shown in Fig.~\ref{fig:setup}(b).
For the transmission measurements, we introduce an open-type waveguide with a width of $122\,\U{mm}$ and height of $25\,\U{mm}$ (see details in Ref.~\cite{Nakanishi2012b}).
A single structure (i.e., meta-atom) is placed at the center of the waveguide.
The meta-atom can effectively interact with the electromagnetic fields confined in the waveguide, whose width is much smaller than the operating wavelength $\lambda \sim 300\,\U{mm}$.
The boundary condition in the experiment differs from that of the simulation described in the previous section.
However, the expected response of the metamaterial is almost the same because the electromagnetic fields around the metamaterial under the waveguide can be regarded as uniform TEM waves.
A probe wave with the frequency $f$
from a network analyzer (Agilent Technologies, E5701C) and a control wave
with the frequency $f\sub{c}$
from a signal generator (Agilent Technologies, N5183A) are combined and fed into the waveguide.
After interacting with the meta-atom in the waveguide, the output wave
is sent to the network analyzer, which can acquire only the frequency
component at $f$. (We ignore the signal at $f=f\sub{c}$.)
The power of the probe wave is kept at $P\sub{probe}=-20\,\U{dBm}$ at the input of the waveguide, and the nonlinear effect induced by the probe wave is negligible.
On the other hand, the power of the control wave $P\sub{c}$ is much greater than  $P\sub{probe}$ to significantly modulate the capacitances of the diodes.

First, we measure the transmission spectra without the control wave $P\sub{c}=0$ for the normal incidence $\theta=0$ and orthogonal incidence $\theta=90^\circ$.
The results are shown in Fig.~\ref{fig:setup}(c).
As expected, we observe two common resonances with broad linewidths in both cases, and a sharp resonance appears only for the orthogonal incidence.
The resonances around $1.05$, $0.65$, and $0.4\,\U{GHz}$ can be identified with those of the radiative, trapped, and control modes, respectively.
Despite the different boundary conditions, the experimental results closely agree with the simulation results.

\begin{figure}[]
 \begin{center}
 \includegraphics[scale=1]{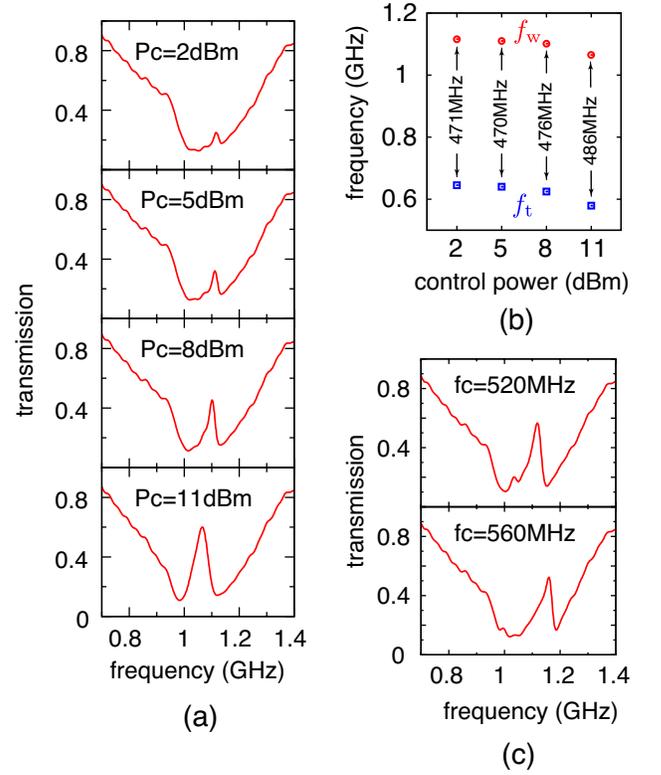}
 \caption{(a) The transmission spectra for $P\sub{c}=2, 5, 8, 11\,\U{dBm}$.
 $f\sub{c}$ is fixed at $480\,\U{MHz}$.
 (b) The resonant frequency of trapped mode $f\sub{t}$ and the position of transparency window $f\sub{w}$.
 (c) The transmission spectra for $f\sub{c}=520$ and $560\,\U{MHz}$.
 $P\sub{c}$ is fixed at $11\,\U{dBm}$. }
 \label{fig:exp}
 \end{center}
\end{figure}

Next, we combine the weak probe wave with the power of $-20\,\U{dBm}$
and the control wave with various powers of $P\sub{c}=2, 5, 8,$ and $11\,\U{dBm}$ to measure the transmission spectra of the probe wave for the normal incidence $\theta=0$.
The frequency of the control wave is $f\sub{c}=480\,\U{MHz}$, which is located at the skirt of the resonance of the control mode, as shown in Fig.~\ref{fig:setup}(c).
We obtaine the transmission spectra for the probe waves from
$0.7$ to $1.4\,\U{GHz}$ as shown in Fig.~\ref{fig:exp}(a).
We can observe a transparent region in a broad resonance dip in each case.
The higher $P\sub{c}$ results in a wider transparency window, as 
explained in Sec.~\ref{sec:theory}.
If there is no need to change $f\sub{c}$, we can increase the efficiency of the control wave by reducing the resistance in the central bar owing to the resonance enhancement of the control wave.

The positions of the transparency windows $f\sub{w}$ and the resonant frequency of the trapped mode $f\sub{t}$ should satisfy the relation $f\sub{w}=f\sub{t}+f\sub{c}$, which provides the two-photon resonance condition.
In the experiment, the resonant frequency of the trapped mode $f\sub{t}$ depends on the power of the control wave $P\sub{c}$ because of the four-wave mixing process \cite{Shadrivov2006,Powell2007,Poutrina2010,Huang2010}, which is ignored in the analysis with the circuit model, as discussed in Sec.~\ref{sec:model}.
To identify $f\sub{t}$ for various $P\sub{c}$, we measure the
transmission spectra for the orthogonal incidence $\theta=90^\circ$ in the presence of the control wave and estimate $f\sub{t}$ from the position of the sharp transmission dip.
The squares in Fig.~\ref{fig:exp}(b) show the estimated $f\sub{t}$.
Clearly, $f\sub{t}$ shifts to a lower frequency with stronger control waves.
The circles in Fig.~\ref{fig:exp}(b) represent the positions of the transparency windows $f\sub{w}$, which are estimated from Fig.~\ref{fig:exp}(a).
As $P\sub{c}$ increases, the transparency windows $f\sub{w}$ shift to a lower frequency in the same manner as $f\sub{t}$.
Consequently, the differences $f\sub{w} - f\sub{t}$, which are also displayed for each $P\sub{c}$ in the graph, stay nearly constant around $480\,\U{MHz}$, and the two-photon resonance condition $f\sub{t}+f\sub{c} \sim f\sub{w}$ is verified to be satisfied.

The two-photon resonance condition $f\sub{t}+f\sub{c} = f\sub{w}$ clearly depends on the frequency of the control wave $f\sub{c}$, and we can control the position of the transparency window by changing $f\sub{c}$.
Figure~\ref{fig:exp}(c) shows the transmission spectra for the normal incidence $\theta=0$ in the presence of the $11$-$\U{dBm}$ control wave with the frequencies of $f\sub{c}=520$ and $560\,\U{MHz}$.
Increasing $f\sub{c}$ causes the transparency window to shift to a higher frequency, as expected.
In addition, the transmission spectra, especially for $560\,\U{MHz}$, exhibit asymmetric shapes, which can be regarded as Fano effects in the metamaterial.
The controllability of $f\sub{w}$ and the spectral shape by external fields is one of the most significant advantages of this EIT metamaterial.

The minimum value of the transparency window is limited by the linewidth
of the trapped mode $\gamma\sub{t}$.
In the ideal case, where the loss of the trapped mode is dominated by the radiation loss, the linewidth of the trapped mode is extremely narrow compared to that of the radiative mode.
This is confirmed in the simulation result shown in Fig.~\ref{fig:simulation}(a), which is obtained without Ohmic losses.
However, in the experiment, the dissipation in the varactor diodes
dominates the loss of the trapped mode, which degrades the performance
of the EIT effect.
The finite $\gamma\sub{t}$ is also responsible for
the great enhancement of the transparency for higher $P\sub{c}$ in the experiment,
because 
the minimum value of the imaginary part of Eq.~(\ref{chi}) for finite $\gamma\sub{t}$
is proportional to $\gamma\sub{t}/(\gamma\sub{t} \gamma +
\Omega\sub{c}^2)$,
which rapidly decreases around $\Omega\sub{c} \sim \sqrt{\gamma\sub{t} \gamma}$.
If we can realize the ideal condition by lowering the Ohmic loss of the nonlinear capacitances, higher transmission in a narrower spectral region can be expected.

\section{Conclusion}

We propose a metamaterial to realize a true EIT effect, where the incidence of the auxiliary electromagnetic wave
induces transparency for the target wave.
The coherence evolution of the quantum EIT system and the time evolution of the charge oscillation in the circuit model of the metamaterial are expressed in the same form, and there is no difference in the susceptibilities derived from the two systems.
This ensures that the proposed metamaterial responds to the probe wave in the same manner as the atomic EIT medium.
In an experiment in the microwave region, we demonstrate that the width
of the transparency window for the probe wave can be controlled by the power of the control wave.
We also show that the position of the transparency peak can be controlled by the frequency of the control wave, and we observe an asymmetric line shape unique to the Fano resonance.

This paper focuses on the case of $\omega\sub{t}+\omega\sub{c} \sim \omega\sub{r}$, but the method can also be adopted for $\omega\sub{r}+\omega\sub{c} \sim \omega\sub{t}$.
In the latter case, the two-photon resonance condition is slightly modified to $\omega=\omega\sub{t}-\omega\sub{c}$, and the increase in the control frequency $\omega\sub{c}$ leads to a decrease in the frequency of the transparency peak.
This case corresponds to a ladder-type system \cite{Xiao1995}, not the
$\Lambda$-type system shown in Fig.~\ref{fig:atom}.

Our method provides
three free parameters in the control wave: amplitude $E$, phase $\phi$, and frequency $\omega\sub{c}$,
while only amplitude is available for static-electric-field-induced transparency.
As demonstrated in the experiment, the position of the transparency
window and the spectral shape can be controlled by $\omega\sub{c}$ without changing the
structure of the metamaterial.
While the transmission spectrum is independent of $\phi$,
the phase of the oscillation in the trapped mode is determined by
$\phi$.
This property is also important in the atomic EIT system,
where the phase of the control light is converted into spin coherence.
For example, it is used to manipulate the direction of a retrieved signal
in a storage and retrieval experiment through the phase-matching condition
\cite{Zibrov2002}.
We expect that most of the applications demonstrated for the atomic EIT system \cite{Fleischhauer2005}, such as the storage of light and enhancement of four-wave mixing, can be realized with the metamaterial.

\acknowledgements

We gratefully thank Yoshiro Urade for his helpful comments.
The present research was supported by JSPS KAKENHI Grants
No. 22109004, No. 25790065, and No. 25287101.

\bibliographystyle{apsrev4-1}
%\bibliography{paper}


\end{document}